\documentclass[sigconf,authordraft, review=false]{acmart}

\AtBeginDocument{%
  \providecommand\BibTeX{{%
    Bib\TeX}}}

\usepackage{bm}
\usepackage{algorithmic}
\usepackage{graphicx}
\usepackage{subcaption}
\usepackage{textcomp}
\usepackage{xcolor}
\usepackage{array}
\usepackage{booktabs}
\usepackage{multirow}
\usepackage{nicefrac}

\setcopyright{acmlicensed}
\copyrightyear{2025}
\acmYear{2025}
\acmDOI{XXXXXXX.XXXXXXX}

\acmConference[Under Review]{Under Review}{July 2024}{}

\def\BibTeX{{\rm B\kern-.05em{\sc i\kern-.025em b}\kern-.08em
    T\kern-.1667em\lower.7ex\hbox{E}\kern-.125emX}}

\begin{document}

\title{Multi-Stream TSN Gate Control Scheduling in the Presence of Clock Synchronization}

\author{Aviroop Ghosh}
\affiliation{%
  \institution{Carleton University}
  \city{Ottawa}
  \country{Canada}}
\email{avighosh@cmail.carleton.ca}

\author{Saleh Yousefi}
\affiliation{%
 \institution{Urmia University}
 \city{Urmia}
 \country{Iran}}
\email{s.yousefi@urmia.ac.ir}

\author{Thomas Kunz}
\affiliation{%
  \institution{Carleton University}
  \city{Ottawa}
  \country{Canada}}
\email{tkunz@sce.carleton.ca}

\begin{abstract}
With the advancement of technologies like Industry 4.0, communication networks must meet stringent requirements of applications demanding deterministic and bounded latencies. The problem is further compounded by the need to periodically synchronize network devices to a common time reference to address clock drifts. Existing solutions often simplify the problem by assuming either perfect synchronization or a worst-case error. Additionally, these approaches delay the scheduling process in network devices until the scheduled frame is guaranteed to have arrived in the device queue, inducing additional delays to the stream. A novel approach that completely avoids queuing delays is proposed, enabling it to meet even the strictest deadline requirement. Furthermore, both approaches can be enhanced by incorporating network-derived time-synchronization information. This is not only convenient for meeting deadline requirements but also improves bandwidth efficiency.

\end{abstract}



\maketitle
\vspace{-0.5em} 
\section{Introduction}
Modern cyber-physical systems aim to provide efficient, flexible and reliable systems by leveraging recent advances in artificial intelligence, cloud computing and networking technologies. These systems require deterministic and low-latencies communication. To achieve this, it is essential to guarantee bounded end-to-end latencies and minimal jitter in data delivery times. Ethernet is such a technology that can meet these requirements in a cost-effective manner. The IEEE Time-Sensitive Networking (TSN) Task Group (TG) is standardizing these enhancements to achieve the necessary performance improvements \cite{cavalcanti_extending_2019, ieee_8021_time-sensitive_2021}.

The IEEE 802.1Qbv standard \cite{ieee_std_8021qbv_ieee_2016}, also known as Time Aware Shaper (TAS), is used for guaranteeing deterministic and bounded latencies for periodic time-sensitive traffic streams. The TAS mechanism achieves this is by enabling specific queue/queues on a egress port of a switch for transmissions while disabling others. The process is managed using Gate Control Lists (GCLs) which are periodic schedules that control the queue opening and closing given the periodicity of time-sensitive streams. This mechanism allows for the time-domain isolation of time-sensitive streams from non-time-sensitive streams. Determinism is ensured by setting these GCLs such that each switch on a given time-sensitive stream route opens its queues for transmission at a predetermined time interval, synchronized with the expected arrival time of the frames. This coordinated timing across the network ensures that time-sensitive data is transmitted with deterministic delay and jitter. However, for the TAS mechanism to work effectively, accurate time-synchronization is required such that all network devices maintain a common clock reference. 

Each device maintains its own internal clock, which, even if initially synchronized with other devices, will begin to drift due to physical properties and external influences affecting its local clock oscillator. This phenomenon is known as clock drift. To mitigate the impacts of clock drift, IEEE has a standard specifying a time-synchronization mechanism, detailed in the IEEE 802.1AS standard \cite{ieee_std_8021as-2020_revision_of_ieee_std_8021as-2011_ieee_2020}. The standard defines a protocol known as generalized Precision Time Protocol (gPTP). A common device in a network, known as the Grandmaster (GM), sends out periodic messages to synchronize all network devices to its interval clock. These messages are passed on hop by hop, where each node uses the information received by the GM to synchronize with its internal clock, accounting for delays on the path such as propagation, processing and transmission delays.     

Deriving GCL schedules for multiple time-sensitive (TS) streams is inherently complex due to varying stream features such as periodicity and frame size, all while meeting specific deadlines. The impact of clock drift adds further complications. To simplify this, existing scheduling approaches either assume perfect synchronization, rendering clock drift impacts inconsequential, or adopt a worst-case error which is a constant factor signifying the maximum possible time difference between any two network devices. The former approach is impractical in realistic networks where clock drift impacts are significant. The latter, as will be demonstrated in this study, introduces unnecessary delays to TS frames. This approach is ineffective for streams having tight jitter requirements such as zero jitter \cite{iic_time_2019}, unless specialised adjustments are made to the GCLs. Furthermore, this study also hypothesises that deriving clock drift information from network measurements related to the gPTP mechanism, and using it for formulating GCLs leads to more efficient schedules.

Considering these, the contributions of this paper are as follows:
\begin{itemize}
    \item \textbf{Alternative Approaches:} Proposing a new scheduling approach for multi-stream GCL scheduling that ensures the minimum possible end-to-end delays while guaranteeing deadline requirements for TS streams.
    \item \textbf{Augmentation with Network Measurements:} Both the existing approach and the proposed approach are augmented by considering network derived measurements in the form of clock drift. 
    \item \textbf{Quantifiable Comparison Method:} A quantifiable comparison method is introduced to compare the different scheduling methods devised. 
    \item \textbf{Elaborated Network Scenarios:} Detailed network scenarios are examined to illustrate how incorporating clock drift measurements optimize bandwidth efficiency.  
\end{itemize}

The rest of the paper is organized as follows: in Section \ref{sec:Literature_Survey}, a survey of the relevant literature is provided. Section \ref{sec:Background} covers the relevant background and system model. Building on these, the scheduling approaches for these studies are developed in Section \ref{sec:Scheduling_approach}. Finally, in Section \ref{sec:Results_Analysis} results and analysis are discussed, and a conclusion is provided in Section \ref{sec:Conclusion}. 

\section{Literature Survey}
\label{sec:Literature_Survey}
This section reviews the influential works in the field to provide relevant context. Note that the emphasis is on offline scheduling algorithms, where all streams are known in advance and a single schedule is generated. Online scheduling algorithms, which handle streams added dynamically have a different evaluation methodology and formulation \cite{stuber_performance_2024}, are outside the scope of this study.  

Many algorithms aim to reduce end-to-end latency for TS streams and prevent frames from different streams to overlap over a given egress port. This is complicated due to non-determinism introduced by clock drift. For this, isolation techniques, such as assigning streams to different egress queues, offsetting transmission start times, and allowing only a single queue to transmit, are used to prevent frame interference.

Isolation techniques are used by Craciunas {et al.} in \cite{craciunas_scheduling_2016}. They use Satisfiability Modulo Theories (SMT) to create GCL schedules. Building on this, Raagaard {et al.} in \cite{raagaard_optimization_2017} use an Integer Linear Programming (ILP) method to develop GCL schedules as presented in their technical report. Barazegaran {et al.} in \cite{barzegaran_real-time_2021} use a constraint programming technique to formulate GCLs. Other studies, such as \cite{vlk_enhancing_2020}, propose switch hardware modifications to ensure detection of frame arrival based on a pre-defined schedule. The authors propose an ILP scheduling approach to determine GCLs and maintain determinism by idling the queue if a frame is missing or delayed.

Some studies focus on jitter control as well. The authors in \cite{durr_no-wait_2016} employ a no-wait approach by avoiding queuing delays entirely. They use an ILP model and tabu search method to generate GCL schedules that ensure the minimum possible latency (no queuing delays) for TS streams, hence no jitter. Other works such as \cite{serna_oliver_ieee_2018,chaine_egress-tt_2022}, use SMT/ILP models for developing GCL schedules while also providing specific provisions for ensuring zero jitter for streams. 

Most studies focus on improving algorithms by striking a balance between finding the optimal solution versus solving times. There are studies which involve other aspects such as deriving GCL schedules after a brief period of loss of synchronization \cite{craciunas_out--sync_2021}. However, all of the studies either consider perfect synchronization or a worst-case error thereby failing to integrate the time synchronization process with TAS. Furthermore, studies considering a worst-case error address the problem by incorporating queuing delays. Specifically, at each switch, the derived GCL will only open a queue to forward time-sensitive traffic once it is confirmed that the frame has been received by considering the worst-case clock drifts between the adjacent devices. Previous research on single TS streams has shown that incorporating network derived measurement, such as clock drift from time synchronization can lead to more bandwidth-efficient schedules \cite{ghosh_importance_2023}. However, no studies have addressed scheduling multiple TS streams in this context. Therefore, this study aims to address this gap. 

\section{Background and System Model}
\label{sec:Background}
This section provides the necessary background beginning with a brief description of the how TSN switches are configured and the topology considered followed by detailing the system model. 

\subsection{Background on TSN Standard}
The TSN standards specify three architectural topologies for configuring TSN switches, namely, fully distributed, hybrid and centralized configuration model. Out of the three, this study considers the centralized configuration model since it is best suited for offline scheduling methods. 

The centralized configuration comprises of the Centralized User Configuration (CUC), which manages user information, and the Centralized Network Controller (CNC), which configures the End Stations (ES) and TSN Switches. The CNC, with a global view of all network devices, communicates with the CUC through the User/Network Interface (UNI). This comprehensive information enables the CNC to devise Gate Control List (GCL) schedules. GCLs, defined for the egress port of a device, operate on periodic cycles, allowing frame transmission when queue gates are open. Frames are assigned to queues based on their Priority Code Point (PCP) markings, and only frames in open gates transmit. If multiple gates are open, the frame in the highest priority queue is transmitted.

\begin{figure}[ht]
\includegraphics[width=8.5cm]{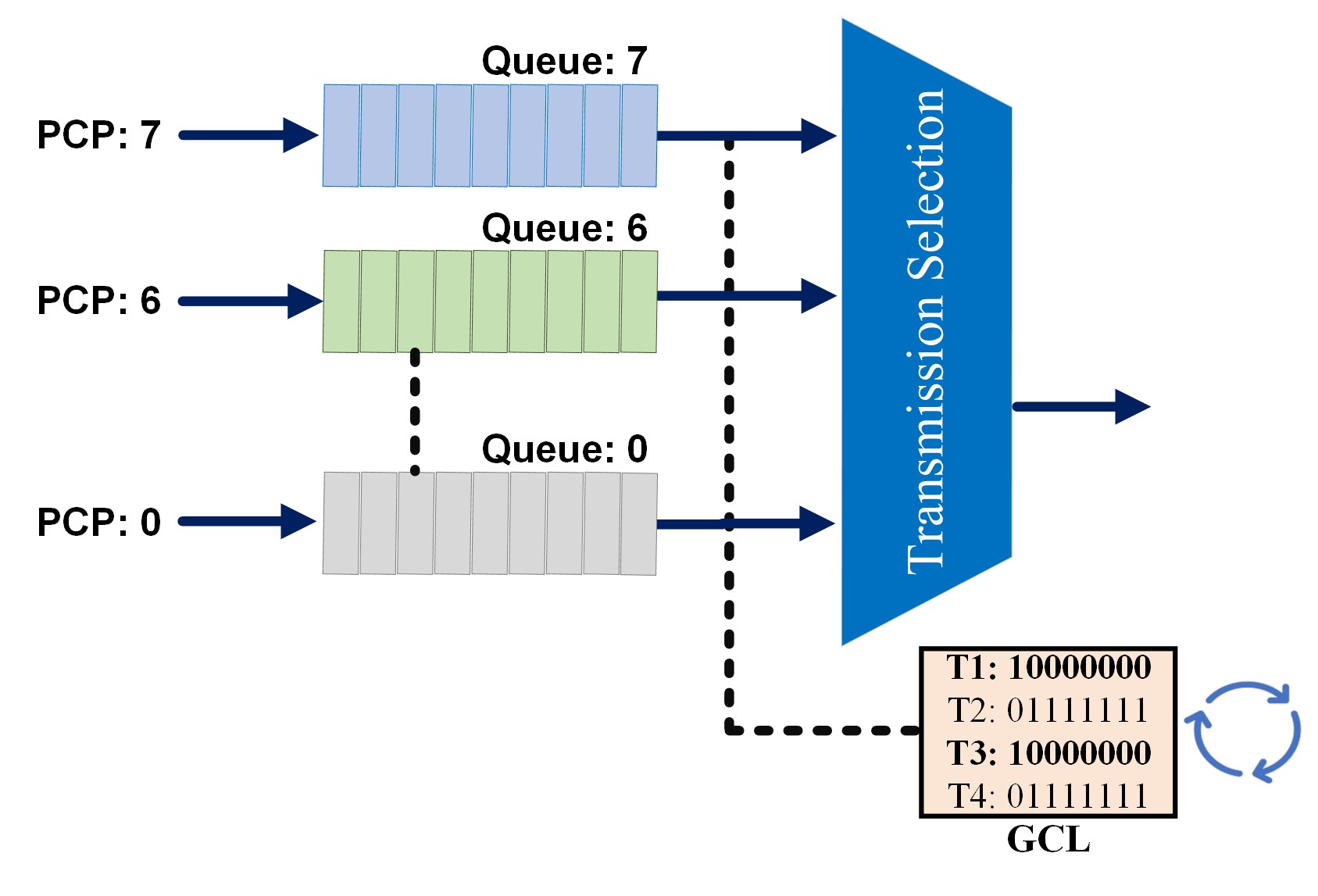}
\centering
\caption{Egress port GCL}
\label{fig:GCL_model}
\end{figure}

Gate open durations for TS traffic are limited, requiring precise scheduling to ensure frames are transmitted within their allocated time slots (scheduling durations). Accurate time synchronization is crucial, as misaligned frame arrivals can cause queuing delays and unnecessary interference. Figure \ref{fig:GCL_model} illustrates the GCL for the egress port of a switch. The schedule repeats every $T$ duration and consisting of sub-periods $T1, T2, T3$ and $T4$. Time-slots $T1$ and $T3$ are allocated for TS traffic, with Queue 7 open, while the other queues are closed (value associated with Queue 7 is set to 1 i.e., ON while others are 0 i.e., OFF). If there is frame misalignment from the previous device during $T1$ and the frame is unable to transmit, then it queues for a duration of at least $T2$ and transmits during $T3$, potentially interfering with the frame scheduled for $T3$.

As explained previously, the worst-case error parameter ($\delta$) used in the literature mitigates these timing inaccuracies. The gPTP synchronization mechanism provides measurements, including the relative clock difference between devices and the GM. These measurements allow for calculating the constant clock drift of a device and establish the relative drift between devices, which should be less than or equal to $\delta$. As an example, consider the worst-case constant clock drift range (measured in parts per million or ppm) as specified by the IEEE 802.1AS standard which is between [$-100$ ppm, $100$ ppm] \cite{ieee_std_8021as-2020_revision_of_ieee_std_8021as-2011_ieee_2020}, i.e., a relative difference of $200$ ppm. With a default synchronization period of $125$ ms, the value of $\delta$ $=$ $25$ $\mu{s}$. However, if the relative difference between any two devices is only $50$ ppm, then the maximum time-difference for the devices will only be $12.5$ $\mu{s}$. Previous studies have shown how the gPTP measurement parameters can be exposed to the TSN control plane \cite{thi_sdn-based_2020}, while the study in \cite{ghosh_importance_2023} showed how to derive the clock drift value from these measurements. 

\subsection{System Model}
To model the network, a directed graph $\mathcal{G} =$ \{$\mathcal{E,V}$\} is considered where $\mathcal{E}$ represents the set of edges that represents the data links and $\mathcal{V}$ represents the set of vertices or the set of devices, namely the End-Stations ($ES$) and Switches ($SW$). A bi-directional, full-duplex link between two consecutive Devices $v_a$ and $v_b$ is denoted by [$v_a$,$v_b$], indicating the direction of flow of traffic for the stream. 

A Stream $s_i$ comprises of periodic data transmission from $ES$ to another $ES$ over a given Route $r_i$. Each Stream $s_i$ is associated with its Payload size $s_i.P$, Periodicity $s_i.T$ and Deadline $s_i.D$. Given that periodic streams can have different periods, they can be aligned into a common periodic network cycle known as a hyperperiod ($hp$). The value is calculated based on the least common multiple (lcm) of the stream periods. Therefore, there can be multiple repetitions of frames from a stream within a $hp$. Consider an example, there are three streams of periodicity $s_i.T = [100, 150, 300]$ $\mu{s}$. The $hp$ $=$ $lcm(100, 150, 300)$ $\mu{s}$ $=$ $300$ $\mu{s}$. Therefore, the first stream has 3 repetitions, the second stream has 2 and the third stream has 1 repeat over the $hp$.   

\begin{figure}[ht]
\includegraphics[width=8.5cm]{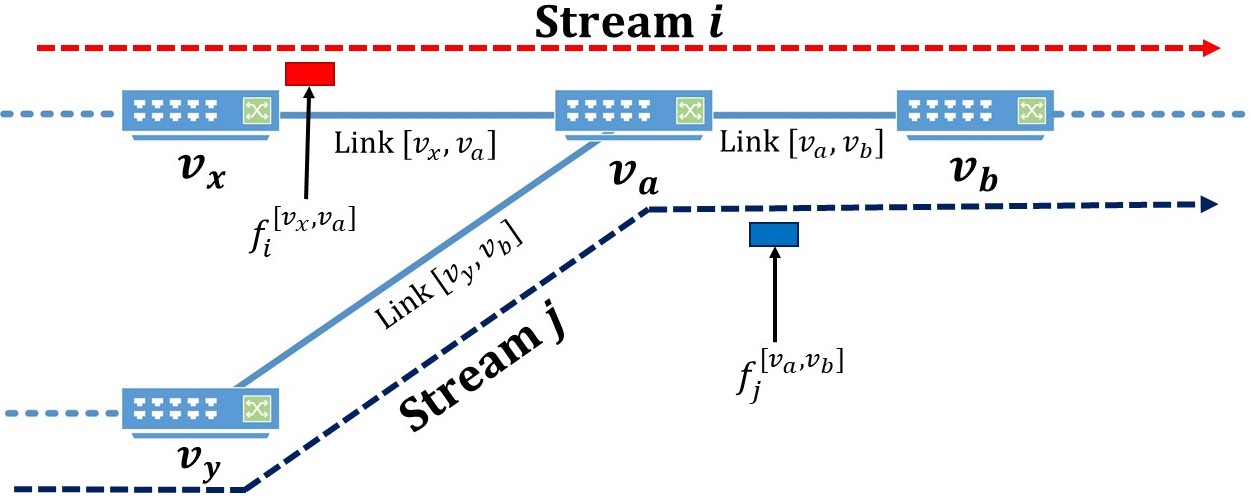}
\centering
\caption{Network model}
\label{fig:Network_model}
\end{figure}

Each stream comprises of a frame transmitted over the route. A frame for the stream is represented as $f_i^{[v_a,v_b]}$. For example, in Fig. \ref{fig:Network_model}, Frame $f_i^{[v_x,v_a]}$ on Link [$v_x$,$v_a$] travels from the egress of Device $v_x$ to the ingress of Device $v_a$, while for Stream $j$, Frame $f_j^{[v_a,v_b]}$ on Link [$v_a$, $v_b$] travels from the egress of Device $v_a$ towards Device $v_b$. The link speed for the link between $v_a$ and $v_b$ is denoted by $[v_a,v_b].c$. Denote the set of routes, streams and frames in the network by $\mathcal{R}$, $\mathcal{S}$ and $\mathcal{F}$ respectively. 

Each frame is associated with a specific set of delays given as:
\begin{itemize}
    \item $[v_a].p$: Processing delay in Device $v_a$.
    \item $[v_a,v_b].d$: Propagation delay on Link [$v_a$,$v_b$].
    \item $f_i^{[v_a,v_b]}.t$: Frame transmission duration of Frame $f_i^{[v_a,v_b]}$ on egress port of Device $v_a$.
    \item $f_i^{[v_a,v_b]}.q$: Queuing delay of Frame $f_i^{[v_a,v_b]}$ in Device $v_a$.
    \item $f_i^{[v_a,v_b]}.L$: Denotes the sum of $f_i^{[v_a,v_b]}.t$, $[v_a,v_b].d$ and $[v_b].p$. 
\end{itemize}

Each frame is transmitted at a specified time based on its GCL schedule. The timing is determined by an offset integer variable $\phi$, where $f_i^{[v_a,v_b]}.\phi$ specifies when the transmission of Frame $f_i^{[v_a,v_b]}$ can start on the egress port of Device $v_a$ (i.e., start of the scheduling duration). These offsets are crucial to control the timing of the gate control operations in the networks and are expressed as integer multiples of macroticks ($mt$). In this study, the value of $mt$ $=$ $0.1$ $\mu{s}$. From the previous example, the $hp$ when expressed in $mt$ is 3000. Now if the offset value of Frame $f_1^{[v_a,v_b]}.\phi$ $=$ $120$ $mt$, this means that the frame is scheduled (i.e., the gate is opened) at $12$ $\mu{s}$ within the $hp$ of 300 $\mu{s}$.

Finally, it is important to discuss the parameters associated with the gPTP mechanism. Each Device $v_a$ has a clock subject to a constant clock drift $v_a.\rho$ and a time-varying clock drift component $v_a.\rho'(t)$. In this study, the time-varying component is neglected ($v_a.\rho'(t)$ $=$ $0$), since it is not considered in the IEEE 802.1AS standard. Given the clock drift and the periodic synchronization (denoted by $T_{s}$), the maximum time difference between two Devices $v_a$ and $v_b$ occurs just before the $T_{sync}$ as the clocks are re-synchronized after this interval. After synchronization, the time difference gradually increases until the next synchronization cycle. For example, assume that $v_a.\rho$ $=$ $10$ ppm and $v_b.\rho$ $=$ $-5$ ppm, with the synchronization interval of $T_{s}$ $=$ $125$ ms, after the initial alignment, the internal clock of $v_a$ will be 1.875 $\mu{s}$ faster than that of $v_b$ after $125$ ms. While formulating GCLs on the basis of gPTP measurements, other factors such as synchronization errors and estimation errors for clock drift could be considered \cite{ghosh_importance_2023}. However, for simplicity in formulation and analysis, these have not been included in this study. 

\subsection{GCL Formation}
In this sub-section, the formation of GCLs based on the offset value is explained briefly. Consider the previous example of three Streams $s_1$, $s_2$ and $s_3$ with periodicity of $[100, 150, 300]$ $\mu{s}$ respectively and $hp$ is $300$ $\mu{s}$. 

\begin{figure}[ht]
\includegraphics[width=8.5cm]{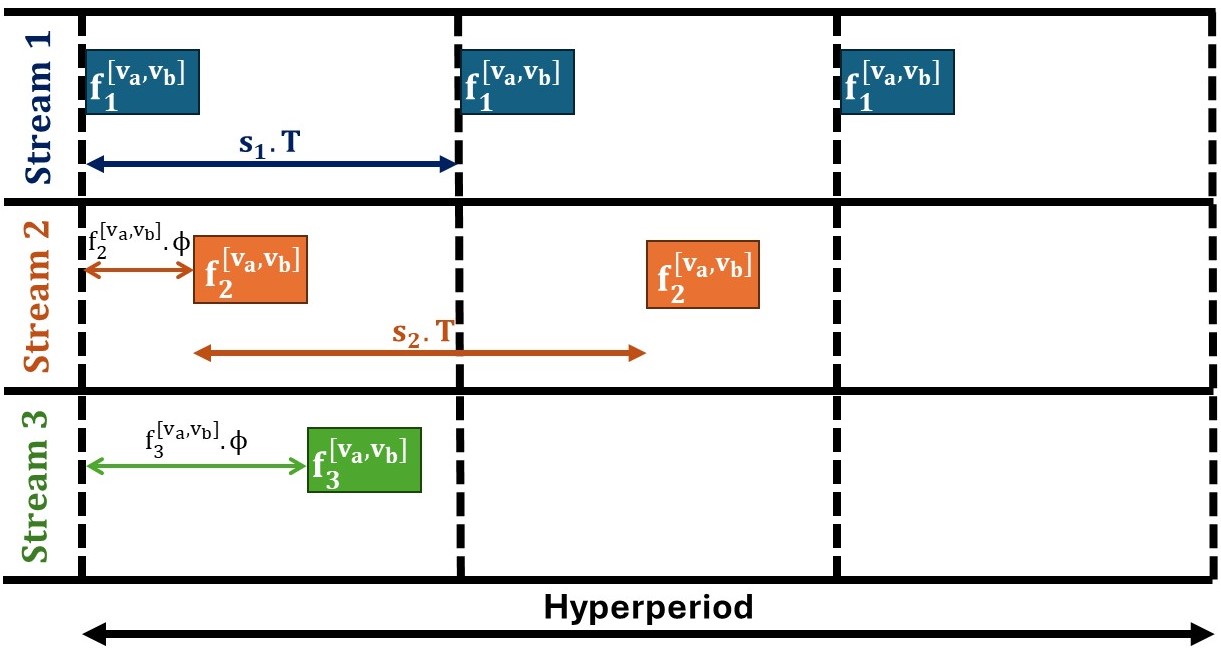}
\centering
\caption{GCL formation example}
\label{fig:hyperperiod}
\end{figure}

An offset integer for a frame in an egress queue $[v_a,v_b]$ specifies the offset within its time period applied consistently across the $hp$. For example, as shown in Fig. \ref{fig:hyperperiod}, if $f_1^{[v_a,v_b]}.\phi$ $=$ $0$, the scheduling duration of Frame $f_1^{[v_a,v_b]}$ starts with the start of the $hp$ and repeats twice with a periodicity of $s_1.T$. Scheduling duration of Frames $f_2^{[v_a,v_b]}$ and $f_3^{[v_a,v_b]}$ are delayed by $f_2^{[v_a,v_b]}.\phi$ and $f_3^{[v_a,v_b]}.\phi$ respectively with Frame $f_2^{[v_a,v_b]}$ occurring once more in the $hp$. 

Thus, based on the offset values and scheduling durations, the GCLs for all egress ports can be calculated, indicating when the queue should be active for transmitting TS frames and when it should be inactive.

\section{Scheduling Approach with Worst-Case Error and Clock Drift}
\label{sec:Scheduling_approach}
In the previous sections the primary motivations of this study were elaborated along with the relevant background and system model description. In this section, this is taken further to define the different scheduling approaches. Before proceeding, it is essential to define the scheduling duration in the context of this study. The scheduling duration for a single TS frame is defined as the transmission duration plus any added slack to mitigate the impacts of uncertainties, such as clock drift. Stating this, there are four approaches considered in this study:
\vspace{-0.5em} 
\begin{itemize}
    \item Worst-Case Delay (WCD) Method: Based on delaying the scheduling duration using a worst-case clock drift estimation ($\delta$). 
    \item Worst-Case Adjustment (WCA) Method: Based on adjusting the scheduling durations to mitigate the impacts of a worst-case clock drift. 
    \item Network-Derived Clock Drift Delay (NCD) Method: Enhances the WCD method by considering network-derived measurements of clock drift. 
    \item Network-Derived Clock Drift Adjustment (NCA) Method: Enhances the WCA method by considering network-derived measurements of clock drift. 
\end{itemize}
\vspace{-0.5em} 
\subsection{Scheduling Duration and Clock Drift}
\label{subsec:scheduling_duration_clock_drift}
Scheduling durations are crucial for forming the GCLs. Inadequate scheduling durations can result in queuing delays, detrimental to TS applications. Conversely, overly generous scheduling durations lead to inefficient bandwidth usage and under-utilization of network resources for other traffic types. Additionally, these challenges are exacerbated by clock drift. To address these issues, two main approaches are used in approaches with respect to scheduling durations: delaying it or adjusting it. This subsection explains the principles behind each approach and the influence of clock drift measurements.

\begin{figure}[ht!]
     \centering
     \begin{subfigure}[b]{0.45\textwidth}
         \centering
         \includegraphics[width=8cm]{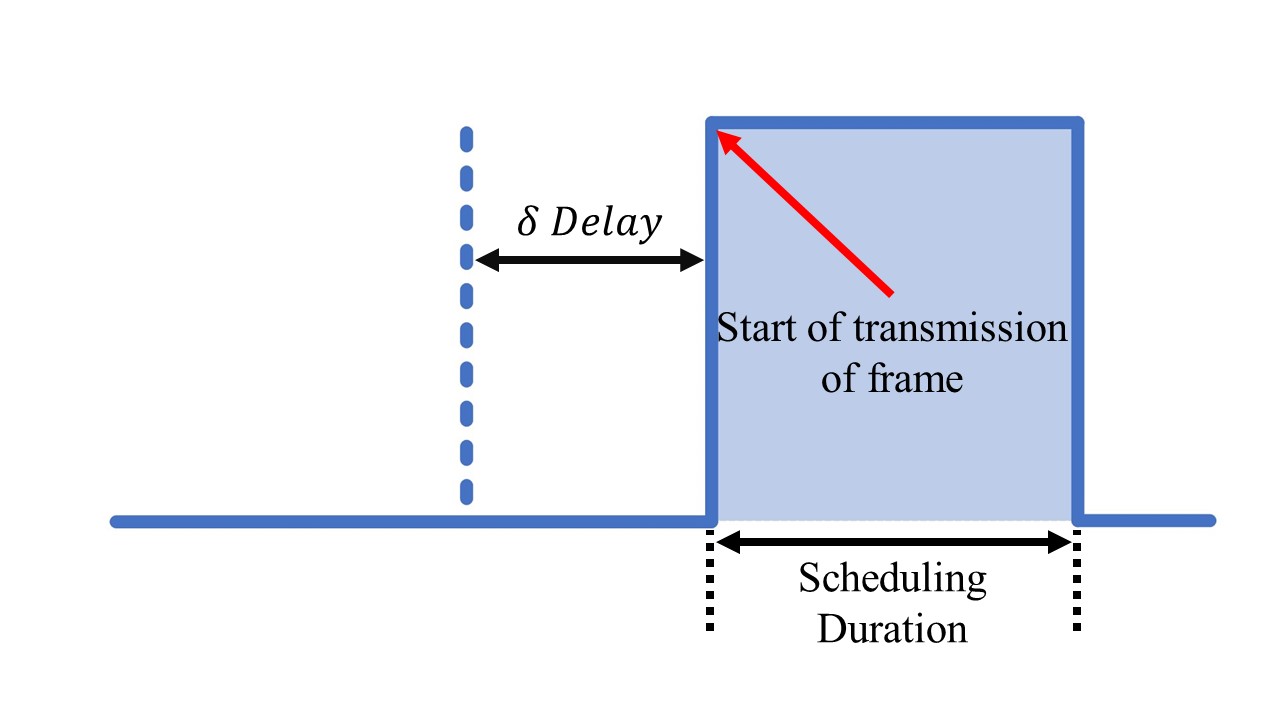}
         \caption{Delaying scheduling duration}
         \label{fig:WCD_SD}
     \end{subfigure}
     \hfill
     \begin{subfigure}[b]{0.45\textwidth}
         \centering
         \includegraphics[width=8cm]{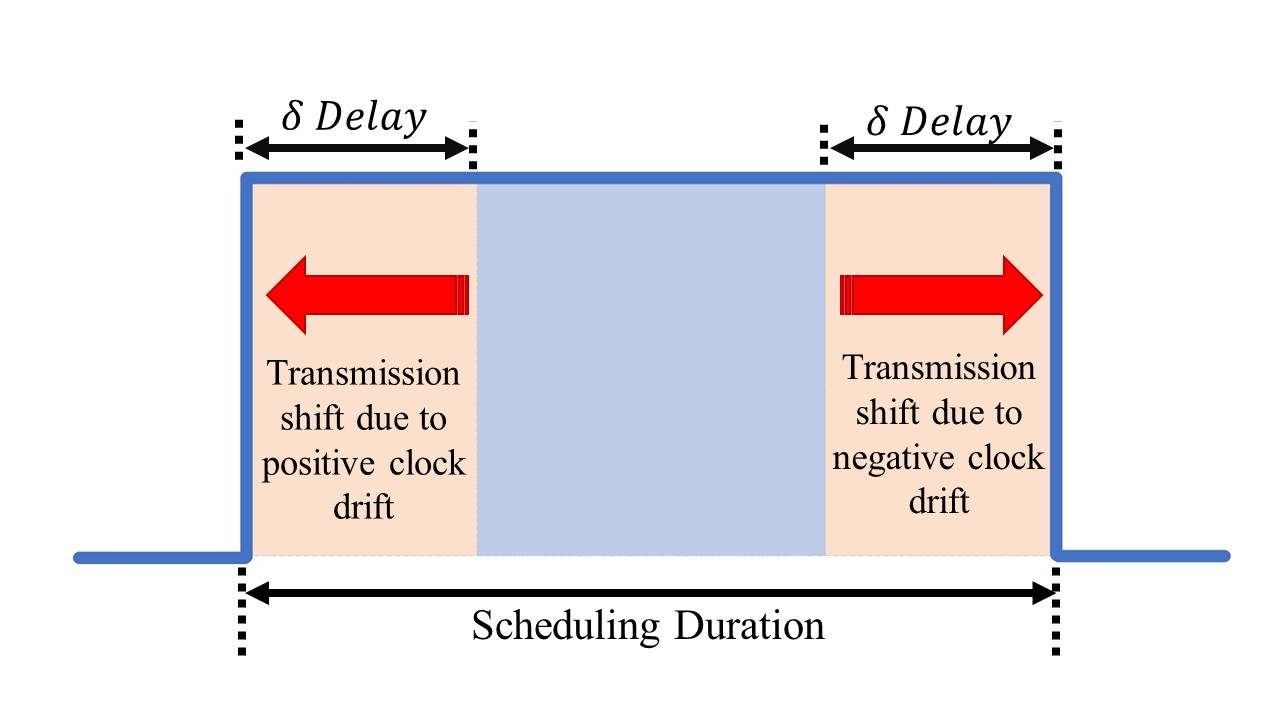}
         \caption{Adjusting scheduling duration}
         \label{fig:WCA_SD}
     \end{subfigure}
     \caption{Illustration of scheduling duration delay and adjustment mechanisms}
     \label{fig:WCA_WCD_SD}
\end{figure}

The WCD scheduling approach, based on works such as \cite{raagaard_optimization_2017, vlk_enhancing_2020,barzegaran_real-time_2021}, uses the scheduling duration delay method. As a TS frame is forwarded through each consecutive switch, the scheduling duration is delayed by a factor of $\delta$. This ensures that the frame is completely received before transmission. Here, the scheduling duration is equal to the transmission duration, resulting in a queuing delay between 0 and $\delta$ for each switch. This is shown in Fig. \ref{fig:WCD_SD}. However, with clock drift considerations, the relative drift between switches can be calculated and adjusted accordingly. For example, if the relative drift between two switches is zero, there is no need for any scheduling duration delay, thereby eliminating any queuing delay. This mechanism is addressed in the NCD approach.

The objective of the WCA method is to match the minimum end-to-end (e2e) latency (therefore, zero jitter). To achieve this, TS frames must be transmitted promptly upon arrival to the queue. Therefore, the scheduling durations are extended to ensure the frames can be transmitted regardless of their arrival times. Therefore, the scheduling duration is adjusted such that both positive and negative clock drifts are considered. This is shown in Fig. \ref{fig:WCA_SD}. With clock drift measurements, these adjustment durations can be shortened and be based on the actual relative drift between devices instead of the worst-case value. This mechanism is captured in the NCA approach.

\subsection{Scheduling Approaches}
In this sub-section, the four scheduling approaches are further elaborated using an Integer Linear Programming (ILP) framework. As detailed in Section \ref{sec:Literature_Survey}, ILP has been frequently used to determine the offsets that govern the GCLs. Furthermore, ILP mechanisms are guaranteed to find an optimal solution as opposed to heuristic solutions \cite{stuber_performance_2024}. Solving the ILP optimization problem yields the offsets ($\phi$) for each stream on each switch, which are then used to create GCL schedules in a post-processing phase. The optimization objective is defined in \eqref{eq:e2e_obj} $\forall s_i \in\mathcal{S}$.
\begin{equation}
    \label{eq:e2e_obj}
    \min[\sum{(\lambda_i - \lambda_i^{min})}]
\end{equation}

The objective minimizes the total excess latency as the difference between the measured e2e latency $\lambda_i$ and the minimum possible e2e latency ($\lambda_i^{min}$) i.e., latency without queuing delays, summed across all streams in the network. While other studies also focus on minimizing the excess number of queues assigned to TS traffic as a part of the optimization problem, this study assumes that only a single queue is dedicated to TS traffic.

Following the definition of the objective function, the constraints applicable to all four approaches are defined. The unit of expression of all the terms are in macroticks. 

\subsubsection{Boundary Conditions:}
All offset values must be greater than or equal to zero. Thus, the lower bound of the offsets for an Stream $i$ for transmission on Link $[v_a,v_b]$ is defined accordingly, as shown in the \eqref{eq:lb}. 
\begin{equation}
    \label{eq:lb}
    f_i^{[v_a,v_b]}.\phi \geq 0 
\end{equation}

The first offset i.e., the offset from the source device of Stream $i$ (denoted by $v_{src\_i}$) is upper-bounded by the hyperperiod as shown in \eqref{eq:ub}
\begin{equation}
    \label{eq:ub}
    f_i^{[v_{src\_i},v_x]}.\phi \leq hp
\end{equation}

Both the relations \eqref{eq:lb} and \eqref{eq:ub} are valid $\forall s_i\in \mathcal{S}$ and $\forall \{[v_a,v_b], \newline [v_{src\_i},v_x]\}\in \mathcal{E}$.

\subsubsection{Non-Overlapping Condition:}
A non-overlapping condition is applied to all the scheduling approaches developed as a part of this study. This condition states that the absolute difference of transmission offsets of any two Streams $i$ and $j$ on Egress Port $[v_a,v_b]$ represented as $f_i^{[v_a,v_b]}.\phi$ and $f_j^{[v_a,v_b]}.\phi$ respectively, must be less than the hyperperiod. This is shown in \eqref{eq:non_overlap}. 
\begin{equation}
    \label{eq:non_overlap}
    \begin{aligned}
    &\forall f_i^{[v_a,v_b]}, f_j^{[v_a,v_b]} \in \mathcal{F}, \forall \{[v_a,v_b]\} \in \mathcal{E}, \forall s_i, s_j \in \mathcal{S}, \; i \neq j: \\
    &|f_i^{[v_a,v_b]}.\phi - f_j^{[v_a,v_b]}.\phi| \le hp - 1
    \end{aligned}
\end{equation}

This prevents any overlaps while deriving GCL schedules. For example, consider $hp$ $=$ $3000$ and $f_i^{[v_a,v_b]}.\phi$ $=$ $0$, if $f_j^{[v_a,v_b]}.\phi$ $=$ $3000$, then this indicates that the offset for Stream $j$ wraps around the GCL period and overlaps with Stream $i$, thereby resulting in an invalid GCL schedule.

\begin{table*}[ht!]
    \centering
    \caption{Set of Expressions}
    \label{tab:set_expressions}
    \begin{tabular}{c m{10cm} c c}
        \toprule
        \textbf{Variable} & \textbf{Expression} & \textbf{Approach} & \textbf{Constraint/Condition} \\
        \midrule
        $\kappa_{11}$ & $\lceil f_i^{[v_x,v_a]}.L + \delta\rceil$ & WCD & Consecutive Offsets \\
        \midrule
        $\kappa_{12}$ & $\lfloor f_i^{[v_x,v_a]}.L + \gamma \times \delta \rfloor$ & WCA & Consecutive Offsets \\
        \midrule
        $\kappa_{13}$ & $\lceil f_i^{[v_x,v_a]}.L + \max\left(|(v_a.\rho - v_x.\rho)T_s|, \psi \times |(v'_a.\rho - v'_x.\rho)T_s|\right)\rceil$ & NCD & Consecutive Offsets \\
        \midrule
        $\kappa_{14}$ & $\begin{aligned}&\lfloor f_i^{[v_x,v_a]}.L + \min\left[0, (v_a.\rho - v_{src\_i}.\rho)T_s, \psi \times (v'_a.\rho - v'_{src\_i}.\rho)T_s\right] \\
        &- \min\left[0, (v_x.\rho - v_{src\_i}.\rho)T_s, \psi \times (v'_x.\rho - v'_{src\_i}.\rho)T_s\right]\rfloor\end{aligned}$ & NCA & Consecutive Offsets \\
        \midrule
        $\kappa_{21}$ & $(\alpha_i \times s_i.T - \beta_j \times s_j.T) - \lceil f_j^{[v_a,v_b]}.t + 1 \rceil$ & WCD, NCD & Scheduling Duration \\
        \midrule
        $\kappa_{22}$ & $(\alpha_i \times s_i.T - \beta_j \times s_j.T) - \lceil f_j^{[v_a,v_b]}.t + 2\delta + 1 \rceil$ & WCA & Scheduling Duration \\
        \midrule
        $\kappa_{23}$ & $\begin{aligned}
        &\left(\alpha_i \times s_i.T - \beta_j \times s_j.T\right) \\ &- \lceil f_j^{[v_a, v_b]}.t + \max\left[0, (v_a.\rho - v_{src\_j}.\rho)T_s, \psi \times (v'_a.\rho - v'_{src\_j}.\rho)T_s\right]  \\
        & - \min\left[0, (v_a.\rho - v_{src\_j}.\rho)T_s, \psi \times (v'_a.\rho - v'_{src\_j}.\rho)T_s\right] + 2\rceil \end{aligned}$ & NCA & Scheduling Duration \\
        \midrule
        $\kappa_{24}$ & $(\beta_j \times s_j.T - \alpha_i \times s_i.T) - \lceil f_i^{[v_a,v_b]}.t + 1 \rceil$ & WCD, NCD & Scheduling Duration \\
        \midrule
        $\kappa_{25}$ & $(\beta_j \times s_j.T - \alpha_i \times s_i.T) - \lceil f_i^{[v_a,v_b]}.t + 2\delta + 1 \rceil$ & WCA & Scheduling Duration \\
        \midrule
        $\kappa_{26}$ & $\begin{aligned}
        &\left(\beta_j \times s_j.T - \alpha_i \times s_i.T\right) \\ &- \lceil f_i^{[v_a, v_b]}.t + \max\left[0, (v_a.\rho - v_{src\_i}.\rho)T_s, \psi \times (v'_a.\rho - v'_{src\_i}.\rho)T_s\right]  \\
        & - \min\left[0, (v_a.\rho - v_{src\_i}.\rho)T_s, \psi \times (v'_a.\rho - v'_{src\_i}.\rho)T_s\right] + 2\rceil \end{aligned}$ & NCA & Scheduling Duration \\
        \midrule
        $\kappa_{31}$ & $(\beta_j \times s_j.T - \alpha_i \times s_i.T) + \lceil f_j^{[v_y,v_a]}.L + \delta \rceil$ & WCD, WCA & Frame Arrival \\
        \midrule
        $\kappa_{32}$ & $(\beta_j \times s_j.T - \alpha_i \times s_i.T) + \lceil f_j^{[v_y,v_a]}.L + \max[0, (v_y.\rho - v_a.\rho)T_s, \psi \times (v'_y.\rho - v'_a.\rho)T_s] \rceil$ & NCD, NCA & Frame Arrival \\
        \midrule
        $\kappa_{33}$ & $(\alpha_i \times s_i.T - \beta_j \times s_j.T) + \lceil f_i^{[v_x,v_a]}.L + \delta \rceil$ & WCD, WCA & Frame Arrival \\
        \midrule
        $\kappa_{34}$ & $(\alpha_i \times s_i.T - \beta_j \times s_j.T) + \lceil f_i^{[v_x,v_a]}.L + \max[0, (v_x.\rho - v_a.\rho)T_s, \psi \times (v'_x.\rho - v'_a.\rho)T_s]\rceil$ & NCD, NCA & Frame Arrival \\
        \midrule
        $\kappa_{41}$ & $f_i^{[v_{n\_i},v_{dst\_i}]}.\phi - f_i^{[v_{src\_i},v_{1\_i}]}.\phi + f_i^{[v_{n\_i},v_{dst\_i}]}.t + f_i^{[v_{n\_i},v_{dst\_i}]}.d$ & WCD, NCD & End-to-End Latency \\
        \midrule
        $\kappa_{42}$ & $f_i^{[v_{n\_i},v_{dst\_i}]}.\phi - f_i^{[v_{src\_i},v_{1\_i}]}.\phi + f_i^{[v_{n\_i},v_{dst\_i}]}.t + f_i^{[v_{n\_i},v_{dst\_i}]}.d + \delta$ & WCA & End-to-End Latency \\
        \midrule
        $\kappa_{43}$ & $\begin{aligned}&f_i^{[v_{n\_i},v_{dst\_i}]}.\phi - f_i^{[v_{src\_i},v_{1\_i}]}.\phi + f_i^{[v_{n\_i},v_{dst\_i}]}.t + f_i^{[v_{n\_i},v_{dst\_i}]}.d \\
        &- \min[0, (v_{n\_i}.\rho - v_{src\_i}.\rho)T_s, \psi \times (v'_{n\_i}.\rho - v'_{src\_i}.\rho)T_s]\end{aligned}$ & NCA & End-to-End Latency \\
        \bottomrule
    \end{tabular}
\end{table*}

\subsubsection{Consecutive Offsets:}
Consecutive offsets ensure that frame transmission on a link starts only after the same frame has been fully transmitted and received by the current device from the previous link (adjusted to clock drift considerations). In the reference topology shown in Fig. \ref{fig:Network_model}, Frame $f_i^{[v_a,v_b]}$ for Stream $i$ can only be transmitted by $v_a$ on Link $[v_a,v_b]$ after Frame $f_i^{[v_x,v_a]}$ is received by Switch $v_a$. The constraint is captured in \eqref{eq:consecutive_offset}. Due to spatial limitations, the equations are expanded in Table \ref{tab:set_expressions} with the $\kappa$ variables capturing the remaining part of the expression.
\begin{equation}
    \label{eq:consecutive_offset}
    \begin{aligned}
    &\forall f_i^{[v_x,v_a]},f_i^{[v_a,v_b]} \in \mathcal{F}, \; \{[v_x,v_a],[v_a,v_b]\} \in \mathcal{E},  \forall s_i \in \mathcal{S}: \\
    &f_i^{[v_a,v_b]}.\phi \geq f_i^{[v_x,v_a]}.\phi + 
    \begin{cases}
    \kappa_{11} & \text{for WCD} \\
    \kappa_{12} & \text{for WCA} \\
    \kappa_{13} & \text{for NCD} \\
    \kappa_{14} & \text{for NCA} \\  
    \end{cases}
    \end{aligned}
\end{equation}

For WCD, the expression dictates that Link $[v_a,v_b]$ can begin transmission only after receiving Frame $f_i^{[v_x,v_a]}$, adjusted for transport delays and the worst-case clock difference between $v_x$ and $v_a$. Similarly, WCA is adjusted with $\delta$, and a term $\gamma$ is included. Here, $\gamma$ $=$ 1 only if the previous Device $v_x$ is an ES; otherwise, $\gamma$ $=$ 0. This adjustment is necessary because the scheduling duration of a switch can be modified to accommodate frame transmission based on the arrival of the frame, while ES transmission time remains fixed, requiring the switch to adjust accordingly.

In the NCD approach, scheduling can begin only after the frame is fully received, accounting for the absolute relative difference in clock drift between devices (considering the polarities of the clock drift values). The consecutive offset for the NCA approach requires that the start of transmission from a device must be at least equal to the start of transmission from the previous device, adjusted for the relative clock drift between the devices and the source ES device ($v_{src\_i}$). The source clock drift serves as a reference point for comparing the relative drifts of $v_x$ and $v_a$.

 Synchronization pathways must also be considered in the NCD and NCA approaches since they are based on network derived clock drift measurements. The term $(v'_x.\rho - v'_{src\_i}.\rho)$ in NCD is introduced to denote whether one device is synchronized with the GM before the other. For example, in Fig. \ref{fig:Network_model}, if $v_y$ is the GM, hop-by-hop synchronization ensures $v_a$ is synchronized before $v_x$. Thus, Frame $f_i^{[v_x,v_a]}$ could be transmitted by $v_x$ while $v_a$ is already synchronized and $v_x$ is not. Here, the clock drift of $v_a$ will be equivalent to the GM ($v_{GM}.\rho$) while $v_x$ remains $v_x.\rho$. Therefore, the relative clock difference is ($v_{GM}.\rho$ $-$ $v_x.\rho$) instead of ($v_{a}.\rho$ $-$ $v_x.\rho$). In this case $v_a'.\rho$ $=$ $v_{GM}.\rho$ while $v_x'.\rho$ $=$ $v_x.\rho$. The variable $\psi = 1$, only if there is a possibility one device is synchronized before another otherwise, $\psi = 0$.

Note that the ceiling function is employed in WCD and NCD methods, whereas the floor function is utilized in WCA and NCA methods. This distinction arises from the nature of the scheduling approaches: the former initiate scheduling duration after full frame reception, while the latter requires it to start just before the frame reaches the queue, reflecting differing principles in the approaches.

\subsubsection{Scheduling Duration Constraint:}
Each link can transmit at most one frame at a time. If there are two frames from different streams, then at any given time, the scheduling duration of frames cannot overlap in time domain. 

In \eqref{eq:scheduling_duration_constraint}, the offset of frames from Streams $i$ and $j$ are set such that they do not interfere when transmitted via Link $[v_a,v_b]$ (with reference to Fig. \ref{fig:Network_model}). Here, $\alpha_i$ and $\beta_j$ represent the number of frame repetitions within the $hp$, such that, $\alpha_i = \{0,...,\nicefrac{lcm(s_i.T,s_j.T)}{s_i.T} - 1\}$ and $\beta_j = \{0,...,\nicefrac{lcm(s_i.T,s_j.T)}{s_j.T} - 1\}$. A logical disjunction is considered such that the scheduling duration of Frame $f_i^{[v_a,v_b]}$ is before that of Frame $f_j^{[v_a,v_b]}$ or after. The source ES device for Streams $i$ and $j$ are denoted by $v_{src\_i}$ and $v_{src\_j}$ respectively.
\begin{equation}
    \label{eq:scheduling_duration_constraint}
    \begin{aligned}
    &\forall f_i^{[v_a,v_b]}, f_j^{[v_a,v_b]} \in \mathcal{F}, \; \forall \{[v_a,v_b]\} \in \mathcal{L}, \\ &\forall s_i, s_j \in \mathcal{S}, \; i \neq j: \\  
    &f_j^{[v_a,v_b]}.\phi - f_i^{[v_a,v_b]}.\phi \leq
    \begin{cases}
    \kappa_{21} & \text{for WCD, NCD} \\
    \kappa_{22} & \text{for WCA} \\ 
    \kappa_{23} & \text{for NCA} \\
    \end{cases} \\
    & \vee \\
    &f_i^{[v_a,v_b]}.\phi - f_j^{[v_a,v_b]}.\phi \leq
    \begin{cases}
    \kappa_{24} & \text{for WCD, NCD} \\
    \kappa_{25} & \text{for WCA} \\
    \kappa_{26} & \text{for NCA} \\
    \end{cases}
    \end{aligned}
\end{equation}

For the WCD and NCD approaches, the scheduling duration is equal to the transmission duration. For the WCA method the scheduling duration adds a factor of $2\delta$ (Section \ref{subsec:scheduling_duration_clock_drift}). In these approaches, a slack of 1 $mt$ is added to ensure that the transmission end of one frame does not overlap with the beginning of another. 

For the NCA method, the start of the scheduling durations of TS frames are adjusted based on the relative drift between the switch and the source, similar to the end of the scheduling duration. Furthermore, this is also adjusted considering the value of $\psi$ where the source or the switch assumes the value of the GM. Therefore, total scheduling duration is the transmission duration plus the maximum adjustment required subtracted with the minimum adjustment required. Furthermore, an adjustment factor of 2 $mts$ is considered. The second $mt$ is introduced to ensure that the scheduling duration encompasses the frame transmission. For example, consider Frame $f_i^{[v_a,v_b]}$, if the frame arrives at $12.09$ $\mu{s}$, then with at $mt$ $=$ $0.1$ $\mu{s}$, the scheduling start $f_i^{[v_a,v_b]}.\phi = 120$ $mts$. If the frame requires a transmission time of $10$ $\mu{s}$ (or $100$ $mts$) i.e., transmits at $22.09$ $\mu{s}$, then the scheduling duration of $101$ $mts$ ensures complete transmission. 

All approaches apply a ceiling function to the scheduling durations to ensure complete frame transmission and total isolation of the frame during its scheduled time.

\subsubsection{Frame Arrival Constraint:}
The non-deterministic effects of clock drift and synchronization errors may cause two frames from different ingress ports to arrive simultaneously at the same egress port. To prevent this, the frame arrival constraint is required. The constraint is shown in \eqref{eq:frame_arrival_constraint}.
\begin{equation}
    \label{eq:frame_arrival_constraint}
    \begin{aligned}
    &\forall s_i, s_j \in \mathcal{S}, \; \forall f_i^{[v_a,v_b]}, f_i^{[v_x,v_a]}, f_j^{[v_y,v_a]} \in \mathcal{F},  \; i \neq j\\
    &\forall \{[v_x,v_a], [v_y,v_a], [v_a,v_b]\} \in \mathcal{E}: \\ 
    &f_i^{[v_a,v_b]}.\phi - f_j^{[v_y,v_a]}.\phi \leq 
    \begin{cases}
    \kappa_{31} & \text{for WCD, WCA} \\
    \kappa_{32} & \text{for NCD, NCA} \\ 
    \end{cases} \\
    & \vee \\
    &f_j^{[v_a,v_b]}.\phi - f_i^{[v_x,v_a]}.\phi \leq 
    \begin{cases}
    \kappa_{33} & \text{for WCD, WCA} \\
    \kappa_{34} & \text{for NCD, NCA} \\
    \end{cases}
    \end{aligned}
\end{equation}

The constraint ensures that, the arrival of Frame $f_j^{[v_y,v_a]}$ to $v_a$ is after the start of scheduling $f_i^{[v_a,v_b]}.\phi$ or that Frame $f_i^{[v_x,v_a]}$ arrives after $f_j^{[v_a,v_b]}.\phi$. The scheduling duration constraint prevents any overlap between the frames. 

Timing mismatches are considered as frames arrive from different devices. For WCD and WCA approaches, a worst-case timing error ($\delta$) is used, representing the maximum possible relative drift between Device $v_x$/$v_y$ and $v_a$. For NCD and NCA approaches, the adjustment is based on the maximum possible relative drift between the devices.

\subsubsection{End-to-End Latency:}
\label{subsec:e2e_latency}
The end-to-end latency equality constraint for Stream $i$ relates the start of the transmission from the source to the first switch (denoted by $v_{1\_i}$) on Route $r_i$, to the transmission onto the final link between the last switch ($v_{n\_i}$) and the destination ES ($v_{dst\_i}$). The e2e latency constraints are listed in \eqref{eq:e2e_latency}.   
\begin{equation}
    \label{eq:e2e_latency}
    \begin{aligned}
    &\forall s_i \in \mathcal{S}, \; \forall f_i^{[v_{src\_i},v_{1\_i}]},f_i^{[v_{n\_i},v_{dst\_i}]} \in \mathcal{F}, \\
    &\forall \{[v_{src\_i},v_{1\_i}],[v_{n\_i},v_{dst\_i}]\} \in \mathcal{E}: \\
    &\lambda_i = 
    \begin{cases}
    \kappa_{41} & \text{for WCD, NCD} \\
    \kappa_{42} & \text{for WCA} \\
    \kappa_{43} & \text{for NCA} \\
    \end{cases}
    \end{aligned}
\end{equation}

Note that in case of the WCA approach, $\delta$ is used to establish the differences in clock between $v_{n\_i}$ and $v_{src\_i}$. While, in the NCA approach if the clock of $v_{n\_i}$ is slower than $v_{src\_i}$, an adjustment is necessary because the scheduling start duration will be earlier than when all devices are synchronized. 

The boundary conditions for the e2e latency ($\lambda_i$) are shown in \eqref{eq:e2e_latency_boundaries_1} for the WCD and NCD approaches, and in \eqref{eq:e2e_latency_boundaries_2} for the WCA and NCA approaches. The latter uses equality between $\lambda_i$ and $\lambda_i^{min}$ since these approaches ensure minimum e2e latency for all frames. The relations are valid $\forall s_i \in \mathcal{S}$.
\begin{equation}
    \label{eq:e2e_latency_boundaries_1}
    \lambda_i^{min} \leq \lambda_i \leq s_i.D  
\end{equation}
\begin{equation}
    \label{eq:e2e_latency_boundaries_2}
    \lambda_i^{min} = \lambda_i \leq s_i.D  
\end{equation}

\section{Results and Analysis}
\label{sec:Results_Analysis}
Before presenting the results and analysis, it is crucial to establish a metric for comparing the different scheduling methods, especially given the two distinct approaches: delay-based and adjustment-based. This is performed in a multi-staged, criteria-based approach: 

\begin{itemize}
    \item \textbf{Criteria 1: Deadlines must be kept} \\
    For the GCL schedule generated for each egress port in all the switches in the network to be valid, the deadlines for all frames from every stream across the network must be met. If any frame misses its deadline, the entire schedule is rendered invalid.
    \item \textbf{Criteria 2: Infeasible schedule} \\
    A schedule is deemed infeasible if the scheduling durations of all the TS streams egressing through an egress port cannot be accommodated within the $hp$. Consider $f_i^{[v_a,v_b]}.SD$ to be the scheduling duration of Frame $f_i^{[v_a,v_b]}$ transmitting on Link $[v_a,v_b]$ from Stream $i$. The relation in \eqref{eq:infeasible_schedule} shows the schedule is infeasible for $v_a$ when the cumulative scheduling durations for all streams (along with their repetitions) is greater than $hp$. Here, $[v_a,v_b].N_s$ represents the number of streams transmitting on Link $[v_a,v_b]$, $\forall s_i \in \mathcal{S}$ and $\forall \{[v_a,v_b]\} \in \mathcal{E}$.
    \begin{equation}
    \label{eq:infeasible_schedule}
        hp <  \sum_{i=1}^{[v_a,v_b].N_s}(\frac{hp}{s_i.T})f_i^{[v_a,v_b]}.SD 
    \end{equation}
    \item \textbf{Criteria 3: Schedulability Cost} \\
     The Schedulability Cost ($SC$) qualifies as the cumulative scheduling durations for all the streams across all the switches encountered over an individual route within the specified $hp$. As shown in \eqref{eq:schedulability_cost}, for each Stream $i$, the total scheduling duration of a frame across Route $r_i$ (denoted by $r_i.SD$) is multiplied by the number of frame occurrences within the $hp$. Summing this parameter across the total number of TS streams in the network (denoted by $N_s$) gives the value of $SC$, $\forall s_i \in \mathcal{S}$ and $\forall r_i \in \mathcal{R}$. 
    \begin{equation}
    \label{eq:schedulability_cost}
    sc = \sum_{i=1}^{N_s}\frac{hp}{s_i.T}\times \frac{r_i.SD}{hp} = \sum_{i=1}^{N_s}\frac{r_i.SD}{s_i.T}
    \end{equation}

    A schedule is stated to be more bandwidth efficient if it has a lower $SC$ than the schedule it is being compared with. 
\end{itemize}

\subsection{Case Studies}
\label{subsec:case_studies}
In this section, a series of case studies are conducted to validate the scheduling approaches and compare the schedules. The approaches are executed using the CPLEX optimization tool to generate the transmission offsets. Subsequently, MATLAB is utilized in a post-processing phase to create the GCL schedules. Finally, the INET framework (version 4.4) for OMNeT++ is employed for network simulation. The simulation period runs for 1 second, allowing sufficient time to analyze the results. 

\begin{figure}[htpb]
\includegraphics[width=8.5cm]{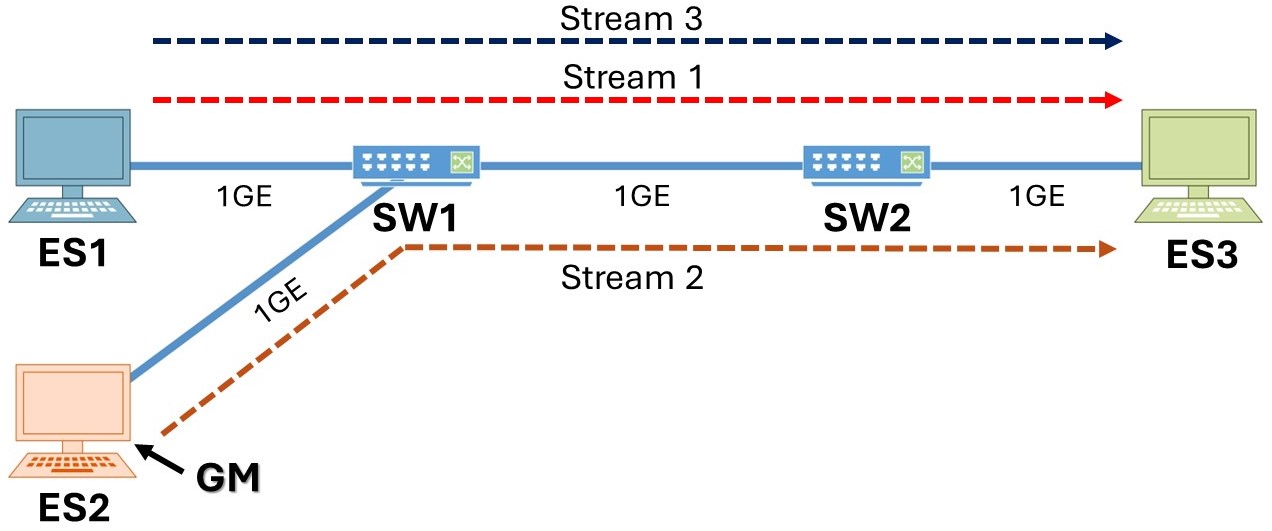}
\centering
\caption{Case study network topology}
\label{fig:case_study}
\end{figure}

First, the network topology is defined as shown in Fig. \ref{fig:case_study}. In this setup, $ES1$ and $ES2$ are the sources, while $ES3$ is the destination. All the links in the network are 1 Gbps Ethernet (GE). There are three traffic streams: $s_1$, $s_2$ and $s_3$. Both $s_1$ and $s_3$ originate from $ES1$ and terminate at $ES3$. Stream $s_2$ originates from $ES2$ and terminates at $ES3$. The routes for the streams are: $r_1 = ([ES1,SW1],[SW1,SW2],[SW2,ES3])$ for $s_1$, $r_2 = ([ES2,SW1],\newline [SW1,SW2],[SW2,ES3])$ for $s_2$ and $r_3 = ([ES1,SW1], \newline [SW1,SW2],[SW2,ES3])$ for $s_3$. Also note, in this topology, $ES2$ is considered the GM.   

Table \ref{tab:system_model} shows the network traffic profiles. The periodicity, deadlines and frame sizes specified for TS traffic are consistent with the requirements specified in \cite{iic_time_2019}. A further assumption is that all the associated delays such as propagation, processing in the network and devices are fixed across the network and the devices. The minimum and maximum clock drift values, represented by $[v_a^{min}].\rho$ and $[v_a^{max}].\rho$ respectively, are chosen specifically for demonstration in this case study. 

The case studies (CS) include three scenarios based on different clock drift values for the devices:

\begin{enumerate}
    \item $[ES1, ES2, SW1, SW2]$ $=$ $[0, 0, +10, -10]$ $ppm$.
    \item $[ES1, ES2, SW1, SW2]$ $=$ $[+10, -10, -10, +10]$ $ppm$.
    \item $[ES1, ES2, SW1, SW2]$ $=$ $[-5, +5, +5, +5]$ $ppm$.
\end{enumerate}

CS1 tests the scheduling approach with a perfect sources (and GM) and extreme clock drift in switches. CS2 explores extreme clock drift in both sources and switches, with interleaving clock drift polarities for devices in routes $r_1$ and $r_3$. CS3 uses lower and closer clock drift values to demonstrate the improvements of NCD and NCA approaches over WCD and WCA respectively, with a polarity change in $ES1$ to prevent uniform clock drift across all devices.

The results presented in Table \ref{tab:case_study} displays the end-to-end latency ranges for each stream across various scenarios and approaches. Furthermore, the $SC$ value is calculated for each scenario based on the specific scheduling approach used.

\begin{table}[ht]
\small
\centering
\caption{System Model Details and Requirements}
\label{tab:system_model}
\begin{tabular}{| m{0.25\linewidth} | m{0.40\linewidth} | m{0.20\linewidth} |}
\hline
\textbf{Parameter} & \textbf{Symbol} & \textbf{Value} \\ \hline
Deadline & $s_i.D$; $\forall s_i \in \mathcal{S}$ & 45.00 $\mu$s \\ \hline
Stream \newline periodicity & $[s_1.T, s_2.T, s_3.T]$ & $[100, 150, 300]$ \newline $\mu$s \\ \hline
Macrotick & $mt$ & 0.1 $\mu$s \\ \hline
Hyperperiod & $hp$ & 3000 \\ \hline
Propagation \newline delay & $[v_a,v_b].d$; \newline $\forall [v_a,v_b] \in \mathcal{E}$ & 50 ns \\ \hline
Processing delay & $[v_a].p$; $\forall [v_a] \in \mathcal{V}$ & 5 $\mu$s \\ \hline
Transmission \newline delay & $f_i^{[v_a,v_b]}.t$; \newline $\forall s_i \in \mathcal{S}$, $\forall [v_a,v_b] \in \mathcal{E}$ & 12.144 $\mu{s}$ \\ \hline
Frame Size & $f_i$; $\forall s_i \in \mathcal{S}$, $\forall f_i \in \mathcal{F}$ & 1464 Bytes \\ \hline
Min e2e latency & $\lambda_i^{min}$; $\forall s_i \in \mathcal{S}$ & 39.682 $\mu$s \\ \hline
Clock drift range & $[v_a^{min}.\rho, \; v_a^{max}.\rho]$; $\forall v_a \in \mathcal{V}$ & $[-10, 10]$ ppm \\ \hline
Synchronization \newline periodicity & $T_{sync}$ & $125$ ms \\ \hline
Worst-case error & $\delta$ & 2.5 $\mu{s}$ \\ \hline
\end{tabular}
\end{table}

\begin{table*}[htpb]
    \centering
    \caption{Case Study Results}
    \label{tab:case_study}
    \begin{tabular}{|c || m{3.2cm} | m{3.2cm} | m{3.2cm} || m{1.5cm} | m{1.5cm} | m{1.5cm}|}
    \hline
    \multirow{2}{*}{\textbf{Method}} & \multicolumn{3}{c||}{\textbf{e2e Latency Range (unit $\bm\mu$s)}} & \multicolumn{3}{c|}{\textbf{Schedulability Cost}}  \\
    \cline{2-7}
    & \textbf{Scenario 1} & \textbf{Scenario 2} & \textbf{Scenario 3} & \textbf{Scenario 1} & \textbf{Scenario 2} & \textbf{Scenario 3} \\ \hline 
    \textbf{WCD} & 
    Stream 1: [44.674, 46.043] \newline Stream 2: [44.734, 46.044] \newline Stream 3: [44.734, 46.043] & 
    Stream 1: [44.673, 44.793] \newline Stream 2: [42.235, 44.792] \newline Stream 3: [44.793, 47.293] & 
    Stream 1: [43.424, 44.793] \newline Stream 2: [44.733, 44.793] \newline Stream 3: [43.543, 44.793] & 
    0.4290 & 0.4290 & 0.4290 \\ \hline
    \textbf{WCA} & 
    Stream 1: [39.682, 39.682] \newline Stream 2: [39.682, 39.682] \newline Stream 3: [39.682, 39.682] & 
    Stream 1: [39.682, 39.682] \newline Stream 2: [39.682, 39.682] \newline Stream 3: [39.682, 39.682] & 
    Stream 1: [39.682, 39.682] \newline Stream 2: [39.682, 39.682] \newline Stream 3: [39.682, 39.682] & 
    0.6920 & 0.6920 & 0.6920 \\ \hline
    \textbf{NCD} & Stream 1: [43.374, 44.743] \newline Stream 2: [43.434, 44.743] \newline Stream 3: [43.394, 44.743] & Stream 1: [44.673, 44.793] \newline Stream 2: [39.735, 42.292] \newline Stream 3: [44.793, 44.793] & Stream 1: [39.682, 40.993] \newline Stream 2: [39.733, 39.793] \newline Stream 3: [39.743, 40.993] & 0.4290 & 0.4290 & 0.4290 \\ \hline
    \textbf{NCA} & Stream 1: [39.682, 39.682] \newline Stream 2: [39.682, 39.682] \newline Stream 3: [39.682, 39.682] & Stream 1: [39.682, 39.682] \newline Stream 2: [39.682, 39.682] \newline Stream 3: [39.682, 39.682] & Stream 1: [39.682, 39.682] \newline Stream 2: [39.682, 39.682] \newline Stream 3: [39.682, 39.682] & 0.5440 & 0.5763 & 0.5280 \\ \hline
    \end{tabular}
\end{table*}

The results show that both the WCA and NCA approaches ensure the minimum end-to-end (e2e) latency for all frames, as evident by the fact that both the minimum and maximum latency values are equal to the minimum possible e2e latency ($\lambda_i^{min}$). The $SC$ for the WCA approach is constant (due to factoring a worst-case error), while the $SC$ parameters for NCA are 16\% to 24\% lower than that of WCA, indicating greater bandwidth-efficiency. The NCA has shorter scheduling durations due to considering relative clock drifts. Even within the NCA results, Scenario 3 provides the lowest $SC$ due to having the smallest relative drift between the devices, while Scenario 2 has the largest $SC$.  

The scheduling durations of the WCD and NCD methods are the same, therefore, the $SC$ for them are equal across all the scenarios. However, factoring clock drift measurements are crucial. Across all the scenarios, the NCD method not only achieves lower relative difference in e2e latency ranges (thereby, lower jitter) but also reduces the maximum e2e latency when compared to the WCD method. Furthermore, the WCD method, due to the delays and considering a worst-case error, violates the deadline requirement (Criteria 1) as observed in Scenarios 1 and 2. Although, the ILP method provides a feasible solution, the deadline constraint for some frames is violated due to clock drift interactions.

Overall, the WCA and the NCA methods guarantee meeting deadlines with their minimum possible e2e latency (and zero jitter) approach. Therefore, they do not face the issue of violating Criteria 1, unlike the WCD and NCD methods. However, the $SC$ is higher because the scheduling durations are longer. Therefore, there could be an increased propensity of violating Criteria 2.

\subsection{Comparing the Scheduling Approaches}
\label{subsec:sc_appraoch_comparison}
In this section, the system model and topology from Section \ref{subsec:case_studies} is extended to compare the WCD and NCD, and WCA and NCA methods. When comparing the WCD and NCD methods, these yield the same $SC$, however, the WCD method introduces greater e2e delays. The difference in the upper-bound of the e2e latencies between the WCD and NCD method will always be equal to $\delta$. This is because the upper-bound considers the extremities of clock drift interleaving, with the source and last switch having the same polarity and intermediate device having the opposite polarity. Consequently, if the maximum queuing time in $SW1$ is just before synchronization, then this correspond to the minimum at $SW2$ and vice-versa, thereby reducing the overall queuing delay. This assumption cannot be extended to the WCD method. Hence there will always be a difference of $\delta$ between the two scheduling approaches. Note, this was observed for Scenario 2, Stream 3, in Tab. \ref{tab:case_study}, where the difference in upper-bound e2e latency is exactly $2.5$ $\mu{s}$.

\begin{figure}[htpb]
\includegraphics[width=8.5cm]{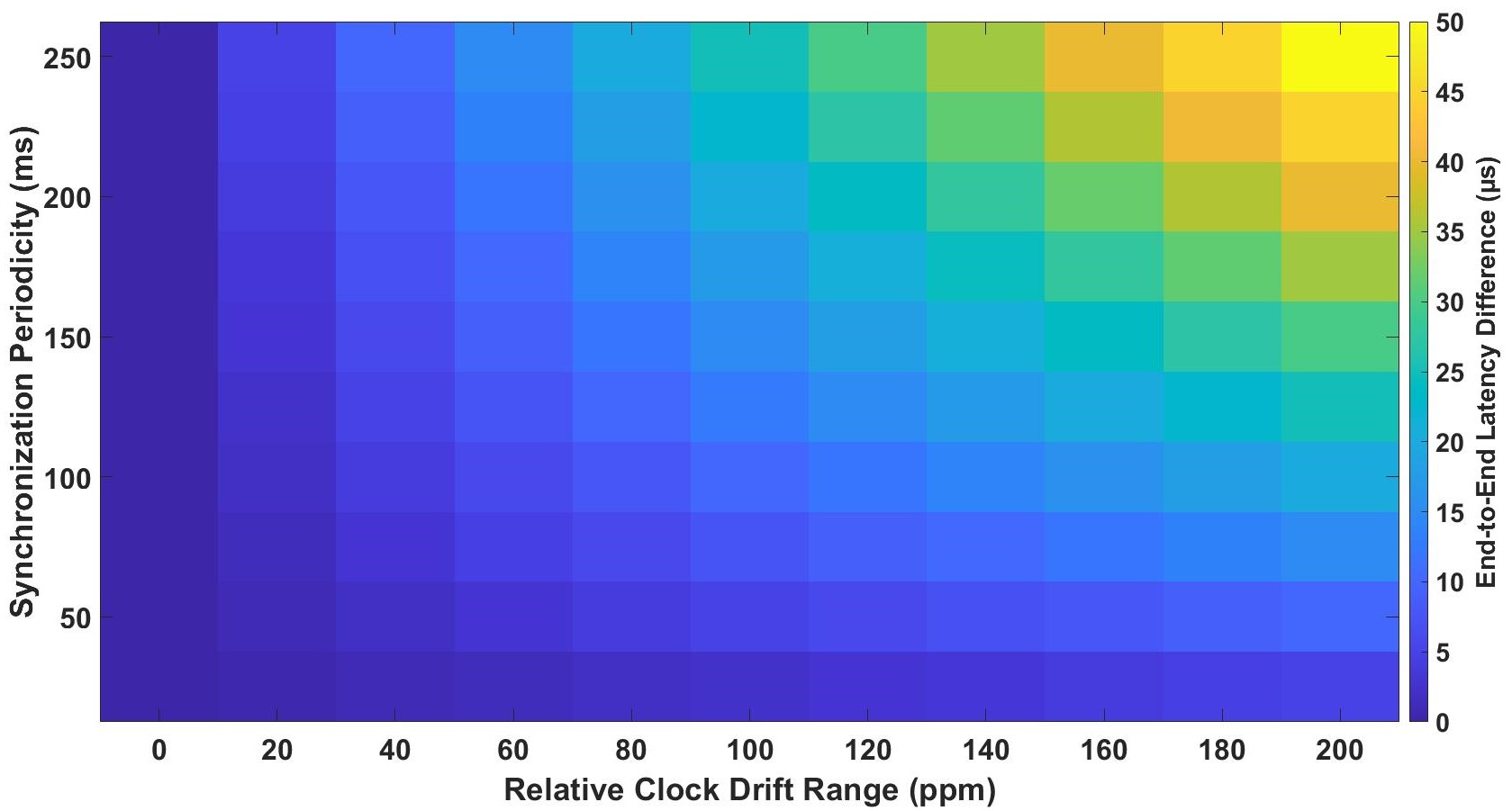}
\centering
\caption{Upper-bound e2e latency difference: WCD and NCD methods}
\label{fig:e2e_latency_diff_WCD_NCD}
\end{figure}

The value of $\delta$ depends on the relative clock drift and synchronization periodicity. As either the clock drift or synchronization periodicity increases, so does $\delta$. Consequently, the upper bound of end-to-end latencies also increases. This is illustrated in Fig. \ref{fig:e2e_latency_diff_WCD_NCD}, where the latency difference ranges from $0$ to $50$ $\mu{s}$, with bluer shades representing a lower difference and the yellow shades a higher difference. The significant extent of this difference is noteworthy. Given that this calculation is for a 3-hop network, any further increase in number of hops further increases the difference. 

\begin{figure}[htpb]
     \centering
     \begin{subfigure}[b]{0.45\textwidth}
         \centering
         \includegraphics[width=8cm]{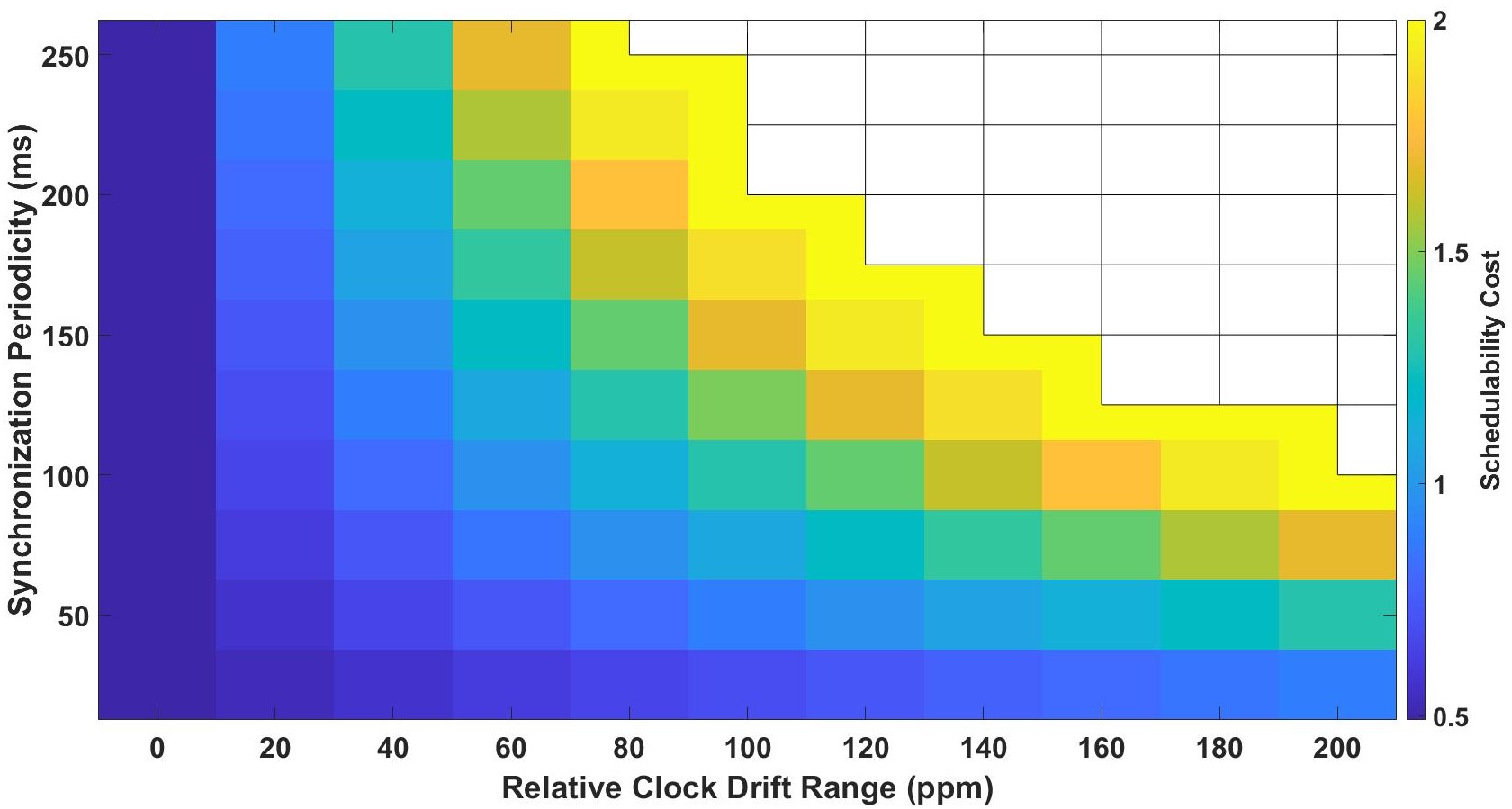}
         \caption{Schedulability cost for the WCA method}
         \label{fig:heatmap_WCA}
     \end{subfigure}
     \hfill
     \begin{subfigure}[b]{0.45\textwidth}
         \centering
         \includegraphics[width=8cm]{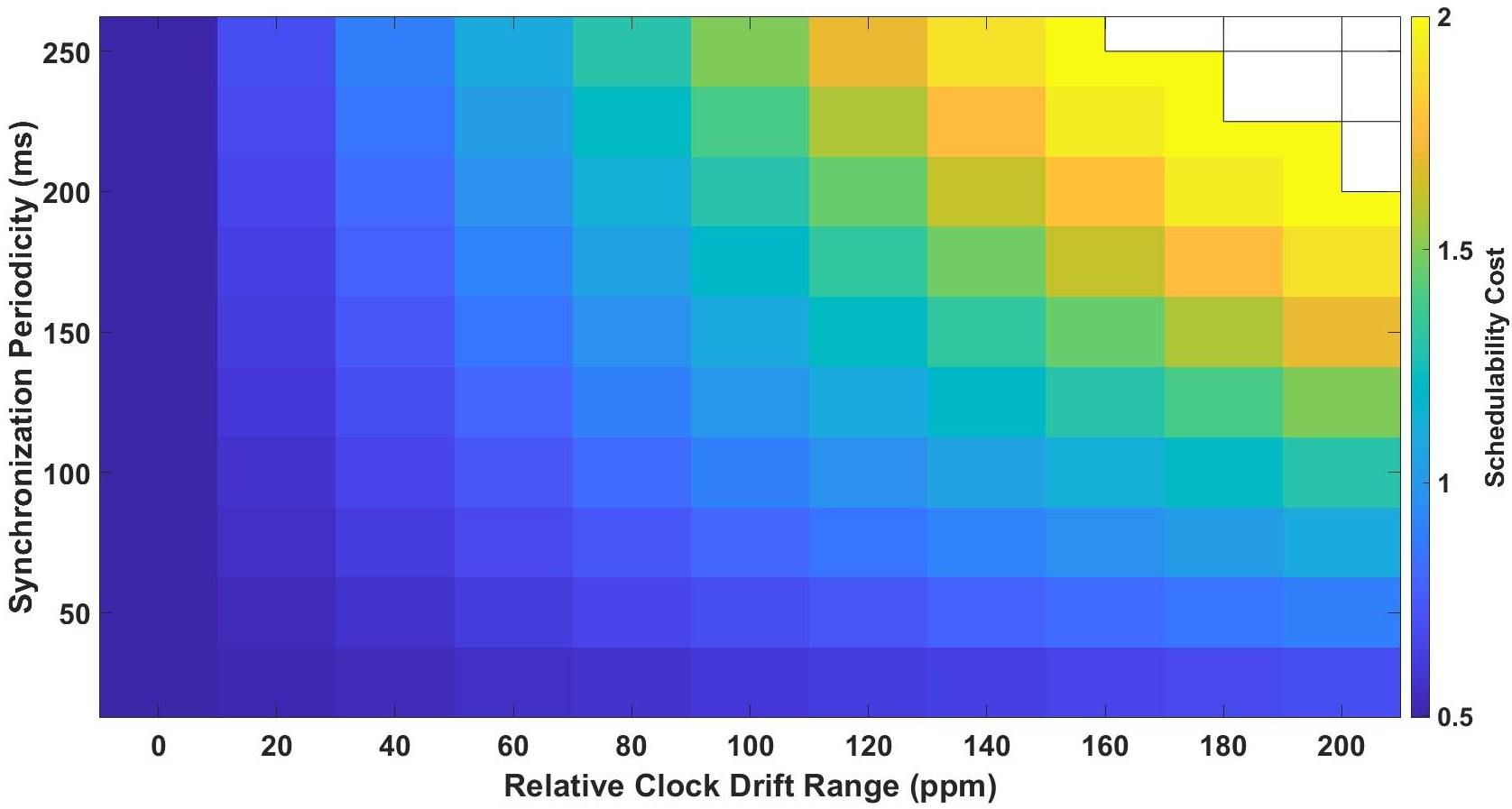}
         \caption{Maximum schedulability cost for the NCA method}
         \label{fig:heatmap_NCA}
     \end{subfigure}
     \caption{Schedulability cost variation for the WCA and NCA methods}
     \label{fig:sc_comparision}
\end{figure}

Conversely, the WCA and NCA methods are designed to achieve the minimum possible e2e latency. Consequently, any improvements by considering clock drift values focus on reducing the $SC$. Using the same topology and system model as in Section \ref{subsec:case_studies}, Fig. \ref{fig:sc_comparision} illustrates how the $SC$ varies with changes in synchronization periodicity and the maximum possible relative clock drift range between two devices.

The $SC$ range for the WCA and NCA methods are shown in Fig. \ref{fig:heatmap_WCA} and Fig. \ref{fig:heatmap_NCA} respectively. The range of values is from $[0.492, 2]$, with bluer shades representing a lower $SC$ and yellow shades representing higher $SC$ values. A point to point comparison shows, the the NCA approach gives lower $SC$ values than the WCA approach. Furthermore, beyond a certain threshold, Criteria 2 is violated for both the methods, as the schedule cannot fit within the hyperperiod. This is depicted by the unshaded (white) region. From the figures, it is clear that the NCA approach has a lower unshaded area than the WCA method due to tighter scheduling durations. This demonstrates that considering clock drift measurements results in more bandwidth-efficient schedules.

\section{Conclusion and Future Work}
\label{sec:Conclusion}
In this study, different scheduling approaches for developing GCLs were derived and compared. Furthermore, a criteria for evaluating GCL schedules including measuring the bandwidth effectiveness was proposed. The scheduling approach primarily used in literature, the WCD approach was augmented by incorporating network derived clock drift measurements, resulting in the NCD approach. As discussed in Section \ref{subsec:sc_appraoch_comparison}, the e2e latency increases as the clocks are synchronized less frequently and/or there is an increase in the number of hops. However, the NCD approach was shown to be more effective at reducing unnecessary delays to the e2e latency and the delay bounds were tighter for TS streams. 

Irrespective of the scheduling approaches, these delay bounds can quickly exceed application deadline requirements based on standardized specifications, necessitating a different approach. This violation of deadlines was especially true of the literature-based WCD approach. Therefore, an alternative and novel scheduling approach was proposed in the form of the WCA which guarantees the minimum e2e latency. However, as discussed in Sections \ref{sec:Results_Analysis}, this results in higher bandwidth usage as indicated by the $SC$ values. By incorporating clock drift measurements and augmenting the approach in the form the NCA method which was demonstrably more bandwidth effective as shown by the $SC$ comparison. 

The feasibility and validity of the approaches, along with the effectiveness of using clock drift measurements, were shown using a series of case studies. Consequently, it can be concluded that the TAS mechanism must be considered in conjunction with time synchronization. Factoring in clock drift measurements leads to more efficient schedules by reducing excessive end-to-end delays or improving bandwidth allocations for TS streams.

As a future work, the approaches will be extended to consider other realistic factors such as measurement errors during the synchronization process and GCL schedule development in complex non-linear topologies. Also, as indicated in Section \ref{subsec:sc_appraoch_comparison}, the synchronization periodicity has an impact on the efficiency of GCL schedules irrespective of the scheduling approaches. Therefore, the schedules can be further optimized by varying the synchronization periodicity. 

\bibliographystyle{ACM-Reference-Format}
\bibliography{references_1.bib}

\end{document}